\documentclass[12pt]{article}
\usepackage{graphicx}
\usepackage[cp1251]{inputenc}
\usepackage{epsfig}
\usepackage[english]{babel}

\date{}

\begin{document}
\setcounter{page}{1}
\pagestyle{plain}

\title{\bf{RKKY interaction in bilayer graphene}}

\author{Yawar Mohammadi$^1$\thanks{Corresponding author. Tel./fax: +98 831 427
4569, Tel: +98 831 427 4569. E-mail address:
yawar.mohammadi@gmail.com} , Rostam Moradian$^{2,3}$}
\maketitle{\centerline{$^1$Young Researchers and Elite Club,
Kermanshah Branch, Islamic Azad University, Kermanshah, Iran}
\maketitle{\centerline{$^2$Department of Physics, Razi University,
Kermanshah, Iran} \maketitle{\centerline{$^3$Nano Science and Nano
Technology Research Center, Razi University, Kermanshah, Iran}

\begin{abstract}

We study the RKKY interaction between two magnetic impurities
located on same layer (intralayer case) or on different layers
(interlayer case) in undoped bilayer graphene in the four-bands
model, by directly calculating the Green functions in the
eigenvalues and eigenvectors representation. Our results show that
both intra- and interlayer RKKY interactions between two magnetic
impurities located on same (opposite) sublattice are always
ferromagnetic (antiferromagnetic). Furthermore we find unusual
long-distance decay of the RKKY interaction in BLG. The intralyer
RKKY interactions between two magnetic impurities located on same
sublattice, $J^{A_{n}A_{n}}(\mathbf{R})$ and
$J^{B_{n}B_{n}}(\mathbf{R)}$, decay closely as $1/R^{6}$ and
$1/R^{2}$ at large impurity distances respectively, but when they
are located on opposite sublattices the RKKY interactions exhibit
$1/R^{4}$ decays approximately. In the interlayer case, the RKKY
interactions between two magnetic impurities located on same
sublattice show a decay close to $1/R^{4}$ at large impurity
distances, but if two magnetic impurities be on opposite
sublattices the RKKY interactions, $J^{A_{1}B_{2}}(\mathbf{R})$
and $J^{B_{1}A_{2}}(\mathbf{R)}$, decay closely as $1/R^{6}$ and
$1/R^{2}$ respectively. Both intra- and interlayer RKKY
interactions have anisotropic oscillatory factors which for
intralayer case is equal to that for single layer graphene. Our
results at weak and strong interlayer coupling limits reduce to
the RKKY interaction of SLG and that of BLG in the two-bands
approximation respectively.

\end{abstract}

%{\it \emph{PACS}}: \emph{74.20.-z, 74.20.Fg}

\vspace{0.5cm}

{\it \emph{Keywords}}: Bilayer graphene; RKKY interaction;
Tight-binding model; Green's function.

%\newpage
\section{Introduction}
\label{sec:1}

Graphene, since its fabrication in 2004\cite{Novoselov1}, has
attracted many experimental and theoretical efforts\cite{Castro
Neto}. These efforts have resulted in discovery of some
unconventional electronic properties such as half-integer quantum
Hall effect\cite{Novoselov2,Gusynin,Peres}, finite conductivity at
zero charge-carrier concentration\cite{Novoselov2,Tworzydlo,Ando},
Klein paradox\cite{Katsnelson1,Katsnelson2}, strong suppression of
weak localization\cite{Morozov}, and Kohn
anomaly\cite{Piscanec,Lazzeri}, etc. These unique properties
originate from two traits of single-layer graphene (SLG), one is
shrinking the Fermi surface to three pairs of points $K$ and
$K^{'}$ ( which are named Dirac points) at the corners of the SLG
first Brillouin zone and the other is linearity of its zero gap
dispersion relation near Dirac points.

The Ruderman-Kittel-Kasuya-Yosida (RKKY) interaction, the indirect
exchange interaction between two magnetic impurities mediated by
the itinerant electrons of the host, is a fundamental quantity of
interest\cite{Ruderman,Kasuya,Yosida,Yafet,Fischer}. Recently,
several
groups\cite{Saremi,Black-Schaffer1,Sherafati,Kogan,Black-Schaffer2,Lee}
have considered the RKKY interaction in SLG, by use of different
methods such as directly computing the Green's function for full
tight-binding band structure\cite{Sherafati}(or for linearly
dispersive band structure\cite{Sherafati,Kogan}) and exact
diagonalization on a finite-size
lattice\cite{Black-Schaffer1,Black-Schaffer2}, and also in
different cases such as presence of the electron-electron
interaction\cite{Black-Schaffer2} or presence of the nonmagnetic
disorders\cite{Lee}. These works have reported two common results:
one is the ferromagnetic (antiferromagnetic) order of magnetic
impurities located on same (apposite) sublattices which originate
from particle-hole symmetry in SLG, or each bipartite lattice,
which is satisfied for nearest-neighbor interactions case. The
other is the long-distance $1/R^{3}$ decay for RKKY interaction in
clean SLG in contrast to the long-distance $1/R^{2}$ decay for the
RKKY interaction in the ordinary two-dimensional metals.
Furthermore the RKKY interaction in undoped and doped single layer
graphene in the presence of a gap are also considered in
Ref.\cite{Roslyak}. This work report $1/R^{3}$ distance dependence
of the RKKY interaction for undoped graphene and mixed $1/R^{2}$
and $1/R^{3}$ distance dependence for doped graphene.

More recently, simultaneous with upward tendency to scrutinize the
properties of SLG, also many researchers have considered the
properties of BLG\cite{Castro Neto}. BLG has a graphene-like Fermi
surface besides a zero gap parabolic dispersion relation near
Dirac points. In spite of these, BLG is very different from
ordinary two-dimensional electron gas (2DEG) with parabolic
spectrum, as well as from SLG. For example, quantum Hall effect,
landau-level degeneracy and Berry
phase\cite{Novoselov2,Zhang,McCann1}, edge states\cite{Castro},
weak localization\cite{Gorbachev}, Coulomb screening and
collective excitations\cite{HWang,XWang} in BLG are different from
those in the ordinary 2DEG and also from those in SLG. The RKKY
interaction in BLG can signalize this difference
more\cite{Jiang,Killi,Parhizgar}. In this paper we study the RKKY
interaction in undoped BLG in the four-band model, by directly
computing the Green's function in eigenvalues and eigenvectors
representation. Also we discussed the RKKY interaction in BLG in
the two limiting cases, weak and strong interlayer coupling. This
paper is organized as follows. In section II, we introduce the
model and obtain the eigenvalues and eigenvectors of BLG. The RKKY
interaction formalism is introduced in section II. In section IV
we present our results for intra- and interlayer RKKY interaction
in BLG. Then we compare our results for the RKKY interaction with
that of SLG and also with that of the ordinary 2DEG. Furthermore
we discuss our results in two limiting cases, weak and strong
interlayer coupling. Finally, we end this paper by summery and
conclusions.

\section{Model}
\label{sec:2}

The BLG (Figure \ref{fig:01} (a)) is composed of a pair of
hexagonal networks of carbon atoms, with $A_{1}$ and $B_{1}$
symmetries on top layer and $A_{2}$ and $B_{2}$ on bottom one.
$A_{1}$ atoms are located above $B_{2}$ atoms and $B_{1}$ atoms
are above the center of hexagons in bottom layer. Let us model the
BLG with only two parameters $t$ and $\gamma$, where $t=2.7eV$ and
$\gamma=0.39eV$ which present the nearest neighbor intralayer
($A_{1}\longleftrightarrow B_{1}$ or $A_{2}\longleftrightarrow
B_{2}$) and interlayer ($A_{1}\longleftrightarrow B_{2}$)
couplings between $\pi$ electrons respectively.

The low energy states of $\pi$ electrons in BLG is given by the
states around the $\mathbf{K}$ and $\mathbf{K}^{'}$ points (Figure
{\ref{fig:01}} (b)) and can be described by a Hamiltonian as

\begin{equation}
\widehat{H}(\mathbf{k})= \left(
\begin{array}{cccc}
0 & \xi v_{F}(k_{x}+\xi ik_{y}) & 0 & \gamma \\
\xi v_{F}(k_{x}-\xi  ik_{y}) & 0 & 0 & 0 \\
0 & 0 & 0 & \xi v_{F}(k_{x}+\xi ik_{y}) \\
\gamma & 0 & \xi v_{F}(k_{x}-\xi  ik_{y}) & 0 \\
\end{array}
\right),\label{eq:01}
\end{equation}
which operates in the space of four components wave functions,
\begin{equation}
\widehat{\Psi}(\mathbf{k})=\left(
\begin{array}{c}
\psi^{A_{1}}(\mathbf{k}) \\
\psi^{B_{1}}(\mathbf{k}) \\
\psi^{A_{2}}(\mathbf{k}) \\
\psi^{B_{2}}(\mathbf{k}) \\
\end{array}
\right),\label{eq:02}
\end{equation}
where $\psi^{\alpha}(\mathbf{k})$ is the electron amplitude on the
sublattice $\alpha$ with $\mathbf{k}$ in the valley
$\mathbf{K}(\mathbf{K}^{'})$. $\mathbf{k}=(k_{x},k_{y})$, two
dimensional momentum, is measured relative to the valley
$\mathbf{K}(\mathbf{K}^{'})$ if $\xi$ be $+1(-1)$ and
$v_{F}=\frac{3}{2}t a$ is the Fermi velocity, where $a$ is the
shortest carbon-carbon distance.

BLG Schrodinger equation,
\begin{equation}
(\widehat{H}(\mathbf{k})-E\mathbf{\widehat{1}})\widehat{\Psi}(\mathbf{k})=0,
\label{eq:03}
\end{equation}
 could be solved to obtain the energy bands
 as\cite{McCann2},
\begin{equation} E_{\lambda
}^{\nu}=\nu(\sqrt{(v_{F}k)^{2}+(\frac{\gamma}{2})^{2}}+(-1)^{\lambda}\frac{\gamma}{2}),
\label{eq:04}
\end{equation}
where $\lambda=1,2$ are the energy bands number and $\nu=+(-)$
indicates the conduction(valance) energy bands respectively. By
using eigenvalues, equation (\ref{eq:04}), we obtain their
appropriate wave functions

\begin{equation}
\widehat{\Psi}_{\mathbf{K}\lambda\nu}(\mathbf{k})=\left(
\begin{array}{c}
\psi^{A_{1}}_{\mathbf{K}\lambda\nu}(\mathbf{k}) \\
\psi^{B_{1}}_{\mathbf{K}\lambda\nu}(\mathbf{k}) \\
\psi^{A_{2}}_{\mathbf{K}\lambda\nu}(\mathbf{k}) \\
\psi^{B_{2}}_{\mathbf{K}\lambda\nu}(\mathbf{k}) \\
\end{array}
\right)=\frac{1}
{\sqrt{2}\sqrt{(E_{\lambda}^{\nu})^{2}+(v_{F}k)^{2}}}\left(
\begin{array}{c}
E_{\lambda}^{\nu}e^{+i{\theta}_{\mathbf{k}}}\\
v_{F}k\\
\nu(-1)^{\lambda}v_{F}k\\
\nu(-1)^{\lambda}E_{\lambda}^{\nu}e^{-i{\theta}_{\mathbf{k}}}
\end{array}
\right),\label{eq:05}
\end{equation}
with $\mathbf{k}$ in the valley $\mathbf{K}$ and
\begin{equation}
\widehat{\Psi}_{\mathbf{K}^{'}\lambda\nu}(\mathbf{k})=\left(
\begin{array}{c}
\psi^{A_{1}}_{\mathbf{K}^{'}\lambda\nu}(\mathbf{k}) \\
\psi^{B_{1}}_{\mathbf{K}^{'}\lambda\nu}(\mathbf{k}) \\
\psi^{A_{2}}_{\mathbf{K}^{'}\lambda\nu}(\mathbf{k}) \\
\psi^{B_{2}}_{\mathbf{K}^{'}\lambda\nu}(\mathbf{k}) \\
\end{array}
\right)=\frac{1}
{\sqrt{2}\sqrt{(E_{\lambda}^{\nu})^{2}+(v_{F}k)^{2}}}\left(
\begin{array}{c}
E_{\lambda}^{\nu}e^{-i{\theta}_{\mathbf{k}}}\\
-v_{F}k\\
-\nu(-1)^{\lambda}v_{F}k\\
\nu(-1)^{\lambda}E_{\lambda}^{\nu}e^{+i{\theta}_{\mathbf{k}}}
\end{array}
\right),\label{eq:06}
\end{equation}
with $\mathbf{k}$ in the valley $\mathbf{K}^{'}$, where
${\theta}_{\mathbf{k}}=tan^{-1}(\frac{k_{y}}{k_{x}})$.

\section{RKKY formalism}
\label{sec:3}

In this section we introduce the RKKY formalism for BLG. Let us
consider two magnetic impurities located on the sublattices
$\alpha$ and $\beta$ in the sites i and j respectively, that have
a small on site spin exchange interaction with the itinerant
electrons as
\begin{equation}
H_{int}=-g\mathbf{S}_{i
\alpha}\mathbf{.}\mathbf{s}_{i\alpha}-g\mathbf{S}_{j\beta}\mathbf{.}
\mathbf{s}_{j\beta},\label{eq:07}
\end{equation}
where $g$ is the coupling constant between the localized magnetic
impurities and the itinerant electrons ($g\ll t$ and $\gamma$),
$\mathbf{S}_{i \alpha}$ is the spin of the magnetic impurity
located at the sublattice $\alpha$ in site i,
$\mathbf{s}_{i\alpha}=\frac{1}{2}c_{i\sigma}^{\alpha\dag}
\mathbf{\sigma}_{\sigma\sigma^{'}}c_{i\sigma^{'}}^{\alpha}$ is the
spin of the itinerant electrons. By use of the perturbation theory
for the thermodynamic potential\cite{Abrikosov}, up to the
second-order of $H_{int}$ and by ignoring all terms proportional
to $\mathbf{S}_{i\alpha}^{2}$ and $\mathbf{S}_{j\beta}^{2}$, one
can obtain following relation
\begin{equation}
H_{RKKY}=-J_{ij}^{\alpha\beta}\mathbf{S}_{i
\alpha}\mathbf{.}\mathbf{S}_{j\beta},\label{eq:08}
\end{equation}
where $J_{ij}^{\alpha\beta}$ is the RKKY interaction between two
magnetic impurities located on sublattices $\alpha$ and $\beta$ in
sites i and j respectively. This is given by\cite{Kogan,Cheianov}
\begin{equation}
J_{ij}^{\alpha\beta}=-\frac{g^{2}}{4}\int_{0}^{\frac{1}{T}}
G^{\alpha\beta}(i,j,\tau)G^{\beta\alpha}(j,i,-\tau),\label{eq:09}
\end{equation}
where $G^{\alpha\beta}(i,j,\tau)$ for BLG could be written in the
eigenvectors and eigenvalues representation as
\begin{eqnarray}
G^{\alpha\beta}(i,j,\tau)&=&\frac{s}{\Omega_{BZ}}\sum_{\lambda,\nu}\int
d^{2}\mathbf{k} [ \nonumber \\
&(&e^{+i(\mathbf{K}+\mathbf{k}).\mathbf{R}_{i}}
\psi_{\mathbf{K}\lambda\nu
}^{\alpha})^{\ast}(e^{+i(\mathbf{K}+\mathbf{k}).\mathbf{R}_{j}}\psi^{\beta}_{\mathbf{K}\lambda\nu
})+(e^{+i(\mathbf{K}^{'}+\mathbf{k}).\mathbf{R}_{i}}
\psi_{\mathbf{K}^{'}\lambda\nu
}^{\alpha})^{\ast}(e^{+i(\mathbf{K}^{'}+\mathbf{k}).\mathbf{R}_{j}}\psi^{\beta}_{\mathbf{K}^{'}\lambda\nu
})] \nonumber \\
& \times & e^{-\xi_{\lambda \nu}(k)\tau}(-(1-n_{F}(\xi_{\lambda
\nu}(k)))\theta(\tau)+n_{F}(\xi_{\lambda
\nu}(k))\theta(-\tau)),\label{eq:10}
\end{eqnarray}
where $\xi_{\lambda\nu}=E_{\lambda}^{\nu}-\mu$,
$n_{F}(\xi_{\lambda\nu})=(1+e^{\beta\xi_{\lambda\nu}})^{-1}$ is
the Fermi distribution function, $s=2$ is the spin degeneracy and
$\Omega_{BZ}=\frac{2(2\pi)^{2}}{3\sqrt{3}a^{2}}$ is the area of
the first Brillouin zone. In the next section, by use of this
formalism, we consider the RKKY interaction in BLG.

\section{Results and discussion}
\label{sec:4}

We apply the introduced RKKY formalism to all possible
configurations of two magnetic impurities located on BLG. These
two magnetic impurities could be on same layer (intralayer
interaction case) or on different layers (interlayer interaction
case). For each case, two impurities could be on same sublattice
or on different sublattices. So overall we have four possible
configurations. In the next subsections for these configurations
we calculate the RKKY interactions from obtained relations of
previous section, equations (\ref{eq:08})-(\ref{eq:10}).

\subsection{Magnetic impurities on same sublattice for intralayer case}
\label{subsec:4.1}
 We now calculate intralayer RKKY
interaction between magnetic impurities located on same sublattice
in the undoped BLG ($\mu=0$) at $T=0$. For this, first we must
obtain the required Green's functions from equations
(\ref{eq:04})-(\ref{eq:06}) and (\ref{eq:09})-(\ref{eq:10}).
Equations (\ref{eq:05}), (\ref{eq:06}) and (\ref{eq:10}) show that
for BLG we have
$G^{A_{1}A_{1}}(i,j,\tau)=G^{B_{2}B_{2}}(i,j,\tau)$ and
$G^{B_{1}B_{1}}(i,j,\tau)=G^{A_{2}A_{2}}(i,j,\tau)$\cite{ZWang}.
So,
$J^{A_{1}A_{1}}(\mathbf{R}_{ij})=J^{B_{2}B_{2}}(\mathbf{R}_{ij})$
and
$J^{B_{1}B_{1}}(\mathbf{R}_{ij})=J^{A_{2}A_{2}}(\mathbf{R}_{ij})$
but $J^{A_{1}A_{1}}(\mathbf{R}_{ij})\neq
J^{B_{1}B_{1}}(\mathbf{R}_{ij})$, hence they must be calculated
separately. The required Green's functions to calculate
$J^{A_{1}A_{1}}(\mathbf{R}_{ij})$ are
\begin{eqnarray}
G^{A_{1}A_{1}}(i,j,\tau)=-\frac{1}{\Omega_{BZ}}
(e^{-i\mathbf{K}.\mathbf{R}_{ij}} +e^{-i\mathbf{K}^{'}.
\mathbf{R}_{ij}}) \int
d^{2}\mathbf{k}e^{-i\mathbf{k}.\mathbf{R}_{ij}}( \nonumber
\\\frac{(E_{1}^{+})^{2}}{(E_{1}^{+})^{2}+(v_{F}k)^{2}}
e^{-E_{1}^{+}\tau}+\frac{(E_{2}^{+})^{2}}{(E_{2}^{+})^{2}+
(v_{F}k)^{2}}e^{-E_{2}^{+}\tau}),\label{eq:11}
\end{eqnarray}
and
\begin{eqnarray}
G^{A_{1}A_{1}}(j,i,-\tau)=\frac{1}{\Omega_{BZ}}
(e^{+i\mathbf{K}.\mathbf{R}_{ij}} +e^{+i\mathbf{K}^{'}.
\mathbf{R}_{ij}}) \int
d^{2}\mathbf{k}e^{+i\mathbf{k}.\mathbf{R}_{ij}} (\nonumber
\\\frac{(E_{1}^{+})^{2}}{(E_{1}^{+})^{2}+(v_{F}k)^{2}}
e^{-E_{1}^{+}\tau}+\frac{(E_{2}^{+})^{2}}{(E_{2}^{+})^{2}+
(v_{F}k)^{2}}e^{-E_{2}^{+}\tau}).\label{eq:12}
\end{eqnarray}
By performing the angle integration of equations (\ref{eq:11}) and
(\ref{eq:12}) we obtain
\begin{eqnarray}
G^{A_{1}A_{1}}(i,j,\tau)=-\frac{2\pi}{\Omega_{BZ}}
(e^{-i\mathbf{K}.\mathbf{R}_{ij}} +e^{-i\mathbf{K}^{'}.
\mathbf{R}_{ij}}) \int_{0}^{\infty} k dk J_{0}(kR) (\nonumber
\\\frac{(E_{1}^{+})^{2}}{(E_{1}^{+})^{2}+(v_{F}k)^{2}}
e^{-E_{1}^{+}\tau}+\frac{(E_{2}^{+})^{2}}{(E_{2}^{+})^{2}+
(v_{F}k)^{2}}e^{-E_{2}^{+}\tau}),\label{eq:13}
\end{eqnarray}
and
\begin{eqnarray}
G^{A_{1}A_{1}}(j,i,-\tau)=\frac{2\pi}{\Omega_{BZ}}
(e^{+i\mathbf{K}.\mathbf{R}_{ij}} +e^{+i\mathbf{K}^{'}.
\mathbf{R}_{ij}}) \int_{0}^{\infty} k dk J_{0}(kR)(\nonumber
\\\frac{(E_{1}^{+})^{2}}{(E_{1}^{+})^{2}+(v_{F}k)^{2}}
e^{-E_{1}^{+}\tau}+\frac{(E_{2}^{+})^{2}}{(E_{2}^{+})^{2}+
(v_{F}k)^{2}}e^{-E_{2}^{+}\tau}),\label{eq:14}
\end{eqnarray}
where $J_{0}$ is Bessel function of the zero order,
$\mathbf{R}_{ij}$ is a vector drawn from site i to site j and
$R=|\mathbf{R}_{ij}|$. If we substitute these equations into
equation (\ref{eq:09}) and replace variables $k$ and $\tau$ with
$x=kR$ and $y=\frac{v_{F}}{R}\tau$ respectively, we obtain
following relation for $J^{A_{1}A_{1}}(\mathbf{R}_{ij})$
\begin{eqnarray}
J^{A_{1}A_{1}}(\mathbf{R}_{ij})=\frac{9g^{2}}{64\pi^{2}
t}\frac{1+cos[(\mathbf{K}-\mathbf{K}^{'}).\mathbf{R}_{ij}]}{(R/a)^{3}}
\int_{0}^{\infty}dy [ \int_{0}^{\infty}xdx J_{0}(x)(\nonumber
\\\frac{\varepsilon_{1}}{\sqrt{x^{2}+(\frac{\gamma}{3t}\frac{R}{a})^{2}}}
e^{-\varepsilon_{1}y}+\frac{\varepsilon_{2}}{\sqrt{x^{2}
+(\frac{\gamma}{3t}\frac{R}{a})^{2}}}e^{-\varepsilon_{2}y})]^{2},\label{eq:15}
\end{eqnarray}
where $\varepsilon_{1}=\sqrt{x^{2}+(\frac{\gamma}{3t}
\frac{R}{a})^{2}} -\frac{\gamma}{3t}\frac{R}{a}$,
$\varepsilon_{2}= \sqrt{x^{2}+(\frac{\gamma}{3t}\frac{R}{a})^{2}}
+\frac{\gamma}{3t}\frac{R}{a}$ and $R$ is the distance between two
magnetic impurities. Also by calculating
$G^{B_{1}B_{1}}(i,j,\tau)$ and $G^{B_{1}B_{1}}(j,i,-\tau)$ and
substituting them into equation (\ref{eq:09}) we obtain following
relation for $J^{B_{1}B_{1}}(\mathbf{R}_{ij})$
\begin{eqnarray}
J^{B_{1}B_{1}}(\mathbf{R}_{ij})=\frac{9g^{2}}{16\pi^{2}
t}\frac{1+cos[(\mathbf{K}-\mathbf{K}^{'}).\mathbf{R}_{ij}]}{(R/a)^{3}}
\int_{0}^{\infty}dy [\int_{0}^{\infty}xdx J_{0}(x)(\nonumber
\\\frac{x^{2}}{\varepsilon_{1}^{2}+x^{2}}
e^{-\varepsilon_{1}y}+\frac{x^{2}}{\varepsilon_{2}^{2}
+x^{2}}e^{-\varepsilon_{2}y})]^{2}.\label{eq:16}
\end{eqnarray}

The result of the integrals in equations (\ref{eq:15}) and
(\ref{eq:16}) are always positive. Therefore the intralayer RKKY
interactions between magnetic impurities located on same
sublattice in BLG are always ferromagnetic. This is similar to the
ferromagnetic coupling between magnetic moments located on same
sublattice in SLG\cite{Saremi,Sherafati,Kogan,Black-Schaffer1}.
Also the intralayer RKKY interactions of BLG,
$J^{A_{1}A_{1}}(\mathbf{R}_{ij})$ and
$J^{B_{1}B_{1}}(\mathbf{R}_{ij})$, show anisotropic
$1+cos[(\mathbf{K}-\mathbf{K}^{'}).\mathbf{R}_{ij}]$-type
oscillations which are similar to that of
SLG\cite{Saremi,Sherafati,Kogan,Black-Schaffer1}. But the $R$
dependent features of the intralayer RKKY interactions of BLG and
that of SLG are different, besides the difference between the $R$
dependence of $J^{A_{1}A_{1}}(\mathbf{R}_{ij})$ and
$J^{B_{1}B_{1}}(\mathbf{R}_{ij})$. The $R$ dependence of the RKKY
interactions of BLG, equations (\ref{eq:15}) and (\ref{eq:16}),
could not be determined analytically. Hence to specify the $R$
dependent behavior of $J^{A_{1}A_{1}}(\mathbf{R}_{ij})$ and
$J^{B_{1}B_{1}}(\mathbf{R}_{ij})$ and compare them with the $R$
dependent behavior of RKKY interaction in SLG, we calculate them
from equations (\ref{eq:15}) and (\ref{eq:16}) numerically and
plot them in units of $J^{AA}(\mathbf{R}_{ij})$ of SLG. Plots of
$J^{A_{1}A_{1}}(\mathbf{R}_{ij})$ and
$J^{B_{1}B_{1}}(\mathbf{R}_{ij})$ in units of
$J^{AA}(\mathbf{R}_{ij})$ have been shown in figures \ref{fig:02}
and \ref{fig:03} respectively. Note that the $R$ dependence of
these plots is independent of direction of
$\mathbf{R}_{ij}$(Figure \ref{fig:01}), since the intralayer RKKY
interactions of BLG and SLG have same oscillatory factor.

Figure \ref{fig:02} shows that the intralayer RKKY interaction
between magnetic impurities located on same sublattice $A_{1}$,
$J^{A_{1}A_{1}}(\mathbf{R}_{ij})$, always is weaker than that of
SLG for all impurity distances
($J^{A_{1}A_{1}}(\mathbf{R}_{ij})<J^{AA}(\mathbf{R}_{ij})$).
Furthermore $J^{A_{1}A_{1}}(\mathbf{R}_{ij})$ at small and large
distances falls off faster than $1/R^{3}$, the power law decay of
RKKY interaction of
SLG\cite{Saremi,Sherafati,Kogan,Black-Schaffer1,Roslyak}, specially at
large impurity distances, $J^{A_{1}A_{1}}(\mathbf{R}_{ij})$ decays
approximately as $1/R^{6}$. This behavior of
$J^{A_{1}A_{1}}(\mathbf{R}_{ij})$ is unlike the long-distances
behavior of the RKKY interaction of SLG, $1/R^{3}$, and also
unlike the long-distances behavior of the RKKY interaction of
ordinary 2DEG, $1/R^{2}$.

From figure \ref{fig:03} we see that at small impurity distances,
$R$, by increasing $R$ first
$J^{B_{1}B_{1}}(\mathbf{R}_{ij})/J_{AA}(\mathbf{R}_{ij})$
increases which means $J^{B_{1}B_{1}}(\mathbf{R}_{ij})$ falls of
smoother than $1/R^{3}$. After approaching to a maximum amount
then it decreases which means that
$J^{B_{1}B_{1}}(\mathbf{R}_{ij})$ decays faster than $1/R^{3}$. By
increasing the impurity distances this behavior repeats again but
with smaller maximum amount. At very large impurity distances,
$J^{B_{1}B{1}}(\mathbf{R}_{ij})$ shows a decay close to $1/R^{2}$
which is similar to the long-distance behavior of the RKKY
interaction in the ordinary 2DEG\cite{Fischer}. Also comparison of
figures \ref{fig:02} and \ref{fig:03} exhibit that for all
impurity distances,
$J^{B_{1}B_{1}}(\mathbf{R}_{ij})>J^{A_{1}A_{1}}(\mathbf{R}_{ij})$.
This could be explain in terms of difference of the local density
of states at $A_{1}$ and $B_{1}$ sublattices. Note that the local
density of states of BLG at sublattice $A_{1}$ is $\frac{S|E|}{\pi
v^{2}_{F}}$ for $|E|>\gamma$ and $\frac{S|E|}{2\pi v^{2}_{F}}$ for
$|E|<\gamma$ while he local density of states of BLG at sublattice
$B_{1}$ is $\frac{S|E|}{\pi v^{2}_{F}}$ for $|E|>\gamma$ and
$\frac{S(|E|+\gamma)}{2\pi v^{2}_{F}}$ for
$|E|<\gamma$\cite{ZWang} where $E$ is energy and $S$ is the area
of unit cell in real space. So the indirect RKKY interaction
mediated by itinerant electrons, which depends on density of
states, behaves as
$J^{B_{1}B_{1}}(\mathbf{R}_{ij})>J^{A_{1}A_{1}}(\mathbf{R}_{ij})$.

Let us now discus our results for the intralayer RKKY interactions
of BLG in the two limiting cases, the weak interlayer coupling
($\gamma\rightarrow0$) and the strong interlayer coupling
($\gamma\rightarrow\infty$). In the first limiting case,
$\gamma\rightarrow0$, we obtain following results for intralayer
RKKY interactions of BLG
\begin{equation}
J^{A_{1}A_{1}}(\mathbf{R}_{ij})=\frac{9g^{2}}{64 \pi^{2}
t}\frac{1+cos[(\mathbf{K}-\mathbf{K}^{'}).\mathbf{R}_{ij}]}{(R/a)^{3}}
\int_{0}^{\infty}dy [ \int_{0}^{\infty}xdx J_{0}(x)(
e^{-xy}+e^{-xy})]^{2},\label{eq:17}
\end{equation}
and
\begin{equation}
J^{B_{1}B_{1}}(\mathbf{R}_{ij})=\frac{9g^{2}}{16\pi^{2}
t}\frac{1+cos[(\mathbf{K}-\mathbf{K}^{'}).\mathbf{R}_{ij}]}{(R/a)^{3}}
\int_{0}^{\infty}dy [\int_{0}^{\infty}xdx J_{0}(x)(
\frac{e^{-xy}}{2}+\frac{e^{-xy}}{2}]^{2},\label{eq:18}
\end{equation}
which after performing the integrals\cite{Kogan,Prudnikov,Integ1}
can be written as
\begin{equation}
J^{A_{1}A_{1}}(\mathbf{R}_{ij})=J^{B_{1}B_{1}}(\mathbf{R}_{ij})=
\frac{9g^{2}}{256\pi t}\frac{1+cos[(\mathbf{K}
-\mathbf{K}^{'}).\mathbf{R}_{ij}]}{(R/a)^{3}},\label{eq:19}
\end{equation}
that is equal to equation 23 of Ref. 20 for the RKKY interaction
between magnetic impurities located on same sublattices in SLG.

In the second limiting case, $\frac{v_{F}k}{\gamma}\rightarrow 0$,
the low energy states of BLG are characterized
by\cite{McCann2,Koshino},
\begin{equation} E_{1
\nu}=\nu\frac{\gamma}{2}((1+\frac{4v^{2}_{F}k^{2}}{
\gamma^{2}})^{\frac{1}{2}}-1) \simeq
\nu\frac{(v_{F}k)^{2}}{\gamma}.\label{eq:20}
\end{equation}
Since the low energy states are localized on $B_{1}$ and $A_{2}$
sites\cite{McCann2}, we only discuss the RKKY interaction between
magnetic impurities located on sublattices $B_{1}$ and $A_{2}$,
namely $J^{B_{1}B_{1}}, J^{A_{2}A_{2}}$, $J^{B_{1}A_{2}}$, and
$J^{A_{2}B_{1}}$. We follow the method introduced in this
subsection to obtain the RKKY interaction, but replace variables
$k$ and $\tau$ with $x=kR$ and $y=\frac{v_{F}^{2}}{\gamma
R^{2}}\tau$ respectively while the terms
$x/(\frac{2\gamma}{3t}\frac{R}{a})$ and
$e^{-\frac{2\gamma}{3t}\frac{R}{a}}$ are ignored. Hence we get
following results for $J^{A_{1}A_{1}}(\mathbf{R}_{ij})$ and
$J^{B_{2}B_{2}}(\mathbf{R}_{ij})$,
\begin{eqnarray}
J^{B_{1}B_{1}}(\mathbf{R}_{ij})=J^{A_{2}A_{2}}(\mathbf{R}_{ij})=\frac{27
m g^{2}a^{4}}{16\pi^{2}
}\frac{1+cos[(\mathbf{K}-\mathbf{K}^{'}).\mathbf{R}_{ij}]}{R^{2}}
\int_{0}^{\infty}dy [\nonumber
 \\ \int_{0}^{\infty}xdx J_{0}(x)
e^{-x^{2}y}]^{2},\label{eq:21}
\end{eqnarray}
where $m=\frac{\gamma}{2v_{F}^{2}}$ is the electron effective
mass\cite{McCann2,Koshino}. The integral can be performed
easily\cite{Kogan,Watson,Integ2} and the equation (\ref{eq:21})
can be written as
\begin{equation}
J^{B_{1}B_{1}}(\mathbf{R}_{ij})=J^{A_{2}A_{2}}(\mathbf{R}_{ij})=\frac{27
m g^{2}a^{4}}{32\pi^{2}
}\frac{1+cos[(\mathbf{K}-\mathbf{K}^{'}).\mathbf{R}_{ij}]}{R^{2}}.\label{eq:22}
\end{equation}
This result is the same as equation 38 of Ref.21 but with an extra
scaling factor $\frac{27}{2}$.

\subsection{Magnetic impurities on different sublattices for intralayer case}
\label{subsec:4.2}

Now we calculate the RKKY interaction,
$J^{\alpha\beta}(\mathbf{R}_{ij})$, when $\alpha$ and $\beta$ are
different sublattices in a same layer. In this case all Green's
functions, $G^{\alpha\beta}(i,j,\tau)$, and therefore all
corresponding intralayer RKKY interactions,
$J^{\alpha\beta}(\mathbf{R}_{ij})$, are equal, so it is enough to
calculate one of them. As an example we calculate
$J^{A_{1}B_{1}}(\mathbf{R}_{ij})$. The required Green's functions
to calculate $J^{A_{1}B_{1}}(\mathbf{R}_{ij})$ are
\begin{eqnarray}
G^{A_{1}B_{1}}(i,j,\tau)=\frac{1}{\Omega_{BZ}} \int
d^{2}\mathbf{k}e^{-i\mathbf{k}.\mathbf{R}_{ij}}
(e^{-i(\mathbf{K}^{'}.\mathbf{R}_{ij}-\theta_{\mathbf{k}})}
-e^{-i(\mathbf{K}^.\mathbf{R}_{ij}+\theta_{\mathbf{k}})})
(\nonumber
\\\frac{v_{F}kE_{1}^{+}}{(E_{1}^{+})^{2}+(v_{F}k)^{2}}
e^{-E_{1}^{+}\tau}+\frac{v_{F}kE_{2}^{+}
}{(E_{2}^{+})^{2}+(v_{F}k)^{2}}e^{-E_{2}^{+}\tau}),\label{eq:23}
\end{eqnarray}
and
\begin{eqnarray}
G^{B_{1}A_{1}}(j,i,-\tau)=\frac{1}{\Omega_{BZ}} \int
d^{2}\mathbf{k}e^{+i\mathbf{k}.\mathbf{R}_{ij}}
(e^{+i(\mathbf{K}^{'}.\mathbf{R}_{ij}-\theta_{\mathbf{k}})}
-e^{+i(\mathbf{K}^.\mathbf{R}_{ij}+\theta_{\mathbf{k}})})
(\nonumber
\\\frac{v_{F}kE_{1}^{+}}{(E_{1}^{+})^{2}+(v_{F}k)^{2}}e^{-E_{1}^{+}\tau}+
\frac{v_{F}kE_{2}^{+}
}{(E_{2}^{+})^{2}+(v_{F}k)^{2}}e^{-E_{2}^{+}\tau}).\label{eq:24}
\end{eqnarray}
After the angle integration we get
\begin{eqnarray}
G^{A_{1}B_{1}}(i,j,\tau)=\frac{2\pi}{\Omega_{BZ}}(
e^{-i(\mathbf{K}^{'}.\mathbf{R}_{ij}-\theta_{\mathbf{R}})}
-e^{-i(\mathbf{K}^. \mathbf{R}_{ij}+\theta_{\mathbf{R}})})
\int_{0}^{\infty} kdkJ_{1}(kR)(\nonumber
\\\frac{v_{F}kE_{1}^{+}
}{(E_{1}^{+})^{2}+(v_{F}k)^{2}}
e^{-E_{1}^{+}\tau}+\frac{v_{F}kE_{2}^{+}
}{(E_{2}^{+})^{2}+(v_{F}k)^{2}} e^{-E_{2}^{+}\tau}),\label{eq:25}
\end{eqnarray}
and
\begin{eqnarray}
G^{B_{1}A_{1}}(j,i,-\tau)=\frac{2\pi}{\Omega_{BZ}}
e^{+i(\mathbf{K}^{'}.\mathbf{R}_{ij}-\theta_{\mathbf{R}})}
-e^{+i(\mathbf{K}^. \mathbf{R}_{ij}+\theta_{\mathbf{R}})})
\int_{0}^{\infty} kdkJ_{1}(kR)(\nonumber
\\\frac{v_{F}kE_{1}^{+}
}{(E_{1}^{+})^{2}+(v_{F}k)^{2}}
e^{-E_{1}^{+}\tau}+\frac{v_{F}kE_{2}^{+}
}{(E_{2}^{+})^{2}+(v_{F}k)^{2}}e^{-E_{2}^{+}\tau}),\label{eq:26}
\end{eqnarray}
 where $J_{1}(x)$ is the Bessel function of first order and
 $\theta_{\mathbf{R}}$ is the angle between
 $\mathbf{K}-\mathbf{K}^{'}$ and $\mathbf{R}_{ij}$. By substituting
these equations into equation (\ref{eq:09}), we obtain
\begin{eqnarray}
J^{A_{1}B_{1}}(\mathbf{R}_{ij})=-\frac{9g^{2}}{64 \pi^{2}
t}\frac{1-cos[(\mathbf{K}-\mathbf{K}^{'}).\mathbf{R}_{ij}
+2\theta_{\mathbf{R}}]}{(R/a)^{3}} \int_{0}^{\infty}dy [
\int_{0}^{\infty}xdxJ_{1}(x)(\nonumber
\\\frac{x}{\sqrt{x^{2}
+(\frac{\gamma}{3t}\frac{R}{a})^{2}}}e^{-\varepsilon_{1}y}
+\frac{x}{\sqrt{x^{2}+(\frac{\gamma}{3t}\frac{R}{a})^{2}}}
e^{-\varepsilon_{2}y})]^{2}.\label{eq:27}
\end{eqnarray}
We see that, similar to SLG, the intralayer RKKY interaction
between magnetic impurities located on different sublattices in
BLG is also negative, which indicate an antiferromagnetic RKKY
interaction between them. Also our result for the intralayer RKKY
interaction has an oscillatory factor,
$1-cos[(\mathbf{K}-\mathbf{K}^{'}).\mathbf{R}_{ij}
+2\theta_{\mathbf{R}}]$, which is equal to that
 of $J^{AB}(\mathbf{R}_{ij})$ in
SLG\cite{Sherafati}(Note that
$J^{A_{1}B_{1}}(\mathbf{R}_{ij})=J^{B_{1}A_{1}}(-\mathbf{R}_{ij}$)).
To specify the $R$ dependent behavior of the intralayer RKKY
interaction in BLG we calculate it from equation (\ref{eq:27})
numerically and plot it in units of $J^{AB}(\mathbf{R}_{ij})$ of
SLG. This has been shown in figure \ref{fig:04}. We see that for
all impurity distances, the RKKY interaction between magnetic
impurities located on different sublattices in BLG is less than
that of SLG. Furthermore figure \ref{fig:04} shows that at large
impurity distances $J^{A_{1}B_{1}}(\mathbf{R}_{ij})$ falls off as
$1/R^{4}$ approximately. This behavior is unlike the
long-distances $1/R^{3}$ decay of the RKKY interaction in
SLG\cite{Saremi,Sherafati,Kogan,Black-Schaffer1,Roslyak} and also
unlike the long-distances $1/R^{2}$ decay of the RKKY interaction
in 2DEG\cite{Fischer}. Similar result for power law decay of
$J^{A_{1}B_{1}}(\mathbf{R}_{ij})$ has been reported in Ref.31.

In the weak interlayer coupling limit, $\gamma\rightarrow0$, we
obtain following result for intralayer RKKY interactions of BLG
\begin{equation}
J^{A_{1}B_{1}}(\mathbf{R}_{ij})=-\frac{9g^{2}}{16\pi^{2}
t}\frac{1-cos[(\mathbf{K}-\mathbf{K}^{'}).\mathbf{R}_{ij}+2\theta_{\mathbf{R}}]}{(R/a)^{3}}
\int_{0}^{\infty}dy [\int_{0}^{\infty}xdx J_{1}(x)(
\frac{e^{-xy}}{2}+\frac{e^{-xy}}{2}]^{2},\label{eq:28}
\end{equation}
which after performing the integrals\cite{Integ1} can be written
as
\begin{equation}
J^{A_{1}B_{1}}(\mathbf{R}_{ij})=-\frac{27g^{2}}{256\pi
t}\frac{1-cos[(\mathbf{K}
-\mathbf{K}^{'}).\mathbf{R}_{ij}+2\theta_{\mathbf{R}}]}{(R/a)^{3}},\label{eq:29}
\end{equation}
that is equal to equation 30 of Ref. 20 for the RKKY interaction
between magnetic impurities located on opposite sublattices in
SLG. Also note that
$J^{A_{1}B_{1}}(\mathbf{R}_{ij})=J^{B_{1}A_{1}}(-\mathbf{R}_{ij})$.

\subsection{Magnetic impurities on same sublattice for interlayer case}
\label{subsec:4.3}

In this subsection we calculate interlayer RKKY interaction
between magnetic impurities located on same sublattice namely
$J^{\alpha_{1}\alpha_{2}}(\mathbf{R}_{ij})$ where $\alpha$ is $A$
or $B$. Similar to the previous subsections, first we calculate
the required Green's functions,
$G^{\alpha_{1}\alpha_{2}}(i,j,\tau)$, where for sublattice
$\alpha=A$ they are
\begin{eqnarray}
G^{A_{1}A_{2}}(i,j,\tau)=-\frac{2\pi}{\Omega_{BZ}}(
e^{-i(\mathbf{K}^{'}.\mathbf{R}_{ij}-\theta_{\mathbf{R}})}
-e^{-i(\mathbf{K}^. \mathbf{R}_{ij}+\theta_{\mathbf{R}})})
\int_{0}^{\infty} kdkJ_{1}(kR)(\nonumber
 \\ \frac{v_{F}kE_{1}^{+}
}{(E_{1}^{+})^{2}+(v_{F}k)^{2}}
e^{-E_{1}^{+}\tau}+\frac{v_{F}kE_{2}^{+}
}{(E_{2}^{+})^{2}+(v_{F}k)^{2}} e^{-E_{2}^{+}\tau}),\label{eq:30}
\end{eqnarray}
and
\begin{eqnarray}
G^{A_{2}A_{1}}(i,j,-\tau)=\frac{2\pi}{\Omega_{BZ}}
(e^{+i(\mathbf{K}^{'}.\mathbf{R}_{ij}-\theta_{\mathbf{R}})}
-e^{+i(\mathbf{K}^. \mathbf{R}_{ij}+\theta_{\mathbf{R}})})
\int_{0}^{\infty} kdkJ_{1}(kR)(\nonumber
 \\ \frac{v_{F}kE_{1}^{+}
}{(E_{1}^{+})^{2}+(v_{F}k)^{2}}
e^{-E_{1}^{+}\tau}+\frac{v_{F}kE_{2}^{+}
}{(E_{2}^{+})^{2}+(v_{F}k)^{2}}e^{-E_{2}^{+}\tau}).\label{eq:31}
\end{eqnarray}
If we substitute equations (\ref{eq:30}) and (\ref{eq:31}) into
equation (\ref{eq:09}), we obtain following relation for
$J^{A_{1}A_{2}}(\mathbf{R})$
\begin{eqnarray}
J^{A_{1}A_{2}}(\mathbf{R}_{ij})=\frac{9g^{2}}{64 \pi^{2}
t}\frac{1-cos[(\mathbf{K}-\mathbf{K}^{'}).\mathbf{R}_{ij}
+2\theta_{\mathbf{R}}]}{(R/a)^{3}} \int_{0}^{\infty}dy [
\int_{0}^{\infty}xdxJ_{1}(x)(\nonumber
 \\ \frac{x}{\sqrt{x^{2}
+(\frac{\gamma}{3t}\frac{R}{a})^{2}}}e^{-\varepsilon_{1}y}
-\frac{x}{\sqrt{x^{2}+(\frac{\gamma}{3t}\frac{R}{a})^{2}}}
e^{-\varepsilon_{2}y})]^{2}.\label{eq:32}
\end{eqnarray}

By a glimpse of equation (\ref{eq:32}) we educe several attractive
results. First, the interlayer RKKY interaction between magnetic
impurities located on same sublattice always is positive. This
implies ferromagnetic order of impurity spins located on same
sublattice in different layers of BLG. Second, the interlayer RKKY
interaction has an anisotropic oscillatory factor,
$1-cos[(\mathbf{K}-\mathbf{K}^{'}).\mathbf{R}_{ij}
+2\theta_{\mathbf{R}}]$. Third, in the limiting case of weak
interlayer coupling, $\gamma\rightarrow0$, the interlayer RKKY
interaction between magnetic impurities located on same sublattice
in BLG tends to zero.

Now we consider the $R$ dependence of the interlayer RKKY
interaction between magnetic impurities located on same
sublattice. Plot of $J^{A_{1}A_{2}}(\mathbf{R}_{ij})$ in units of
$J^{AA}(\mathbf{R}_{ij})$ as a function of $R/a$ for
$\mathbf{R}_{ij}$ along the armchair direction (Figure
\ref{fig:01}) has been shown in figure \ref{fig:05}. Note that the
oscillatory factor of $J^{A_{1}A_{2}}(\mathbf{R}_{ij})$,
$1-cos[(\mathbf{K}-\mathbf{K}^{'}).\mathbf{R}_{ij}
+2\theta_{\mathbf{R}}]$, and that of $J^{AA}(\mathbf{R}_{ij})$,
$1+cos[(\mathbf{K}-\mathbf{K}^{'}).\mathbf{R}_{ij}]$, are equal
for $\mathbf{R}_{ij}$ along the armchair direction. Figure
\ref{fig:05} shows that for all impurity distances,
$J^{A_{1}A_{2}}(\mathbf{R}_{ij})<J^{AA}(\mathbf{R}_{ij})$. Also we
see that by increasing the impurity distance, at first
$J^{A_{1}A_{2}}(\mathbf{R}_{ij})/J^{AA}(\mathbf{R}_{ij})$
increases which means that $J^{A_{1}A_{2}}(\mathbf{R}_{ij})$
decays smoother than $1/R^{3}$. Then after approaching to a
maximum amount it decreases and at very large distances it shows
asymptotically $1/R^{4}$ decay, which is unlike the long-distances
behavior of the RKKY interaction in both
SLG\cite{Saremi,Sherafati,Kogan,Black-Schaffer1} and ordinary
2DEG\cite{Fischer}. Long-distances behavior of
$J^{A_{1}A_{2}}(\mathbf{R}_{ij})$ is similar to that of
$J^{A_{1}B_{1}}(\mathbf{R}_{ij})$.

\subsection{Magnetic impurities on different sublattices for interlayer case}
\label{subsec:4.4}

 We now study the interlayer RKKY interaction between
magnetic impurities located on apposite sublattices. Note that two
possible interlayer RKKY interactions,
$J^{A_{1}B_{2}}(\mathbf{R}_{ij})$ and
$J^{B_{1}A_{2}}(\mathbf{R}_{ij})$, are different and should be
calculated separately. After calculating the required Green
functions and substituting them into equation (\ref{eq:09}), we
get following relations for $J^{A_{1}B_{2}}(\mathbf{R}_{ij})$ and
$J^{B_{1}A_{2}}(\mathbf{R}_{ij})$
\begin{eqnarray}
J^{A_{1}B_{2}}(\mathbf{R}_{ij})=-\frac{9g^{2}}{64\pi^{2}
t}\frac{1+cos[(\mathbf{K}-\mathbf{K}^{'}).\mathbf{R}_{ij}+4\theta_{\mathbf{R}}]}{(R/a)^{3}}
\int_{0}^{\infty}dy [ \int_{0}^{\infty}xdx J_{2}(x)(\nonumber
 \\ \frac{\varepsilon_{1}}{\sqrt{x^{2}+(\frac{\gamma}{3t}\frac{R}{a})^{2}}}
e^{-\varepsilon_{1}\tau}-\frac{\varepsilon_{2}}{\sqrt{x^{2}
+(\frac{\gamma}{3t}\frac{R}{a})^{2}}}e^{-\varepsilon_{2}\tau})]^{2},\label{eq:33}
\end{eqnarray}
and
\begin{eqnarray}
J^{B_{1}A_{2}}(\mathbf{R}_{ij})=-\frac{9g^{2}}{16\pi^{2}
t}\frac{1+cos[(\mathbf{K}-\mathbf{K}^{'}).\mathbf{R}_{ij}]}{(R/a)^{3}}
\int_{0}^{\infty}dy [\int_{0}^{\infty}xdx J_{0}(x)(\nonumber
 \\ \frac{x^{2}}{\varepsilon_{1}^{2}+x^{2}}
e^{-\varepsilon_{1}\tau}-\frac{x^{2}}{\varepsilon_{2}^{2}
+x^{2}}e^{-\varepsilon_{2}\tau})]^{2},\label{eq:34}
\end{eqnarray}
where $J_{2}(x)$ is Bessel function of second order. Equations
(\ref{eq:33}) and (\ref{eq:34})
 show that the interlayer RKKY interactions between magnetic impurities on
different sublattices, $J^{A_{1}B_{2}}(\mathbf{R}_{ij})$ and
$J^{B_{1}A_{2}}(\mathbf{R}_{ij})$, are always negative which
indicate antiferromagnetic order. Also
$J^{B_{1}A_{2}}(\mathbf{R}_{ij})$ has an anisotropic
$1+cos[(\mathbf{K}-\mathbf{K}^{'}).\mathbf{R}_{ij}]$ oscillatory
factor which is similar to that of $J^{AA}(\mathbf{R}_{ij})$ in
SLG\cite{Saremi,Sherafati,Kogan,Black-Schaffer1}, but
$J^{A_{1}B_{2}}(\mathbf{R}_{ij})$ has different oscillatory
factor,
$1+cos[(\mathbf{K}-\mathbf{K}^{'}).\mathbf{R}_{ij}+4\theta_{\mathbf{R}}]$.

Now we consider the $R$ dependence of the interlayer RKKY
interaction between magnetic impurities located on apposite
sublattices. Plots of $J^{A_{1}B_{2}}(\mathbf{R}_{ij})$ and
$J^{B_{1}A_{2}}(\mathbf{R}_{ij})$ in units of
$J^{AA}(\mathbf{R}_{ij})$ as a function of $R/a$ for
$\mathbf{R}_{ij}$ along the armchair direction have been shown in
figures \ref{fig:06} and \ref{fig:07} respectively. Figure
\ref{fig:06} shows that for small impurity distances,
$J^{A_{1}B_{2}}(\mathbf{R}_{ij})$ falls off smoother than
$1/R^{3}$ decay of the RKKY interaction in SLG. But by increasing
impurity distance after approaching to a maximum amount,
$J^{A_{1}B_{2}}(\mathbf{R}_{ij})/J^{AA}(\mathbf{R}_{ij})$
decreases which means that $J^{A_{1}B_{2}}(\mathbf{R}_{ij})$
decays faster than RKKY interaction in SLG. At large impurity
distances, $J^{A_{1}B_{2}}(\mathbf{R}_{ij})$ exhibits closely
$1/R^{6}$ decay which is similar to the long distances power law
decay of $J^{A_{1}A_{1}}(\mathbf{R}_{ij})$ in BLG but is unlike
the long distances behavior of the RKKY interaction in both
SLG\cite{Saremi,Sherafati,Kogan,Black-Schaffer1} and ordinary
2DEG\cite{Fischer}.

From figure \ref{fig:07} wee see that for small distances, by
increasing $R$ first
$J^{B_{1}A_{2}}(\mathbf{R}_{ij})/J^{AA}(\mathbf{R}_{ij})$
increases (which means that $J^{B_{1}A_{2}}(\mathbf{R}_{ij})$
falls of smoother than $1/R^{3}$) and after approaching to a
maximum amount then it decreases (which indicates a power law
decay faster than $1/R^{3}$ for
$J^{B_{1}A_{2}}(\mathbf{R}_{ij})$). This behavior repeats at
larger impurity distances again but with smaller maximum amount.
At very large impurity distances $J^{B_{1}A_{2}}(\mathbf{R}_{ij})$
shows a decay close to $1/R^{2}$ which is similar to the long
distance behavior of the RKKY interaction in the ordinary 2DEG.

As we mentioned in the subsection \ref{subsec:4.1} , in the limit
of strong interlayer coupling the low energy states are localized
on $B_{1}$ and $A_{2}$ sublattices, so we only calculate and
discuss the $J^{B_{1}A_{2}}(\mathbf{R}_{ij})$. In this limiting
case we found
\begin{equation}
J^{B_{1}A_{2}}(\mathbf{R}_{ij})=-\frac{27 m g^{2}a^{4}}{32\pi^{2}
}\frac{1+cos[(\mathbf{K}-\mathbf{K}^{'}).\mathbf{R}_{ij}]}{R^{2}}
\int_{0}^{\infty}dy [\int_{0}^{\infty}xdx J_{0}(x)
e^{-x^{2}y}]^{2},\label{eq:35}
\end{equation}
which after performing the integral numerically it can be written
as
\begin{equation}
J^{B_{1}A_{2}}(\mathbf{R}_{ij})=-\frac{27 m g^{2}a^{4}}{32\pi^{2}
}\frac{1+cos[(\mathbf{K}-\mathbf{K}^{'}).\mathbf{R}_{ij}]}{R^{2}}.
\label{eq:36}
\end{equation}
We see that in the strong interlayer coupling limit, the long
distance behavior of the interlayer RKKY interaction in BLG is
similar to that in the ordinary 2DEG but with a different
$1+cos[(\mathbf{K}-\mathbf{K}^{'}).\mathbf{R}_{ij}]$ oscillatory
factor.

\section{Summary and conclusion}
\label{sec:5}

In summary, we studied the intralayer and interlayer RKKY
interaction between two magnetic impurities in BLG in the
four-band model. We derived them semi-analytically by calculating
the Green functions in the eigenvalues and eigenvectors
representation. For this we first obtained the eigenvalues and
eigenvectors of BLG in the four-band model and by using them the
required Green functions for four possible configurations of two
magnetic impurities are calculated. Then we used them to calculate
corresponding intralayer and interlayer RKKY interactions.

First we summarize our results for the intralayer RKKY
interaction. We found that, similar to the RKKY interaction in
SLG, the intralayer RKKY interactions between magnetic impurities
located on same (opposite) sublattices in BLG are always
ferromagnetic (antiferromagnetic). Also the intralayer RKKY
interactions in BLG have
$1+cos[(\mathbf{K}-\mathbf{K}^{'}).\mathbf{R}_{ij}+\theta]$
oscillatory part which is equal to that of the RKKY interaction in
SLG. The phase factor $\theta$ is equal to $0$ when two magnetic
impurities are on same sublattice but it is equal to
$2\theta_{\mathbf{R}}$ when two magnetic impurities are on
apposite sublattices. Furthermore we found that at large impurity
distances, $J^{A_{1}A_{1}}(\mathbf{R}_{ij})$ falls off as
$1/R^{6}$ approximately while $J^{A_{1}B_{1}}(\mathbf{R}_{ij})$
shows closely $1/R^{4}$ decay, which both power law decays are
unlike $1/R^{3}$ decay of the RKKY interaction in SLG, and also
unlike the $1/R^{2}$ decay of the RKKY interaction in ordinary
2DEG. But $J^{B_{1}B_{1}}(\mathbf{R}_{ij})$, similar to ordinary
two dimensional electron gas, at large impurity distances decays
closely as $1/R^{2}$. The other attractive result of our work is
breaking symmetry of the intralayer RKKY between magnetic moments
on same sublattice in BLG. In contrast to SLG which exhibits
$J^{AA}(\mathbf{R}_{ij})=J^{BB}(\mathbf{R}_{ij})$ we found that
$J^{B_{1}B_{1}}(\mathbf{R}_{ij})$ is always stronger than
$J^{A_{1}A_{1}}(\mathbf{R}_{ij})$. This is due to the difference
between the local density of states of BLG at sublattice $B_{1}$
and $A_{1}$. The tunneling between $A_{1}$ and $B_{2}$ leads to
localization of low energy states on $B_{1}$ sublattices
\cite{McCann2}. This means density of low energy states at $B_{1}$
is more than $A_{1}$ \cite{ZWang} and $A_{1}$ and $B_{1}$ are not
equivalent. Hence $A_{1}A_{1}$ and $B_{1}B_{1}$ RKKY couplings are
different (breaking symmetry of RKKY coupling between magnetic
moment on same sublattices). Furthermore we found that for all
impurity distances,
$J^{A_{1}A_{1}}(\mathbf{R}_{ij})<J^{AA}(\mathbf{R}_{ij})$ and
$J^{A_{1}B_{1}}(\mathbf{R}_{ij}) <J^{AB}(\mathbf{R}_{ij})$, but
for $J^{B_{1}B_{1}}(\mathbf{R}_{ij})$ we obtained $J^{B_{1}B_{1}}
(\mathbf{R}_{ij})>J^{BB}(\mathbf{R}_{ij})$ for small impurity
distances and
$J^{B_{1}B_{1}}(\mathbf{R}_{ij})<J^{BB}(\mathbf{R}_{ij})$ for
large impurity distances. It is noticeable that our results for
power law decay of intralayer RKKY couplings are in agreement with
what has been reported in Ref. 31.

Now we present our results for interlayer RKKY interactions in
BLG. We found that, similar to the intralayer case, the interlayer
RKKY interaction between magnetic impurity located on same
(apposite) sublattices are also ferromagnetic (antiferromagnetic).
The interlayer RKKY interactions have
$1+cos[(\mathbf{K}-\mathbf{K}^{'}).\mathbf{R}_{ij}+\theta]$
oscillatory factor, where $\theta$ for
$J^{A_{1}A_{2}}(\mathbf{R}_{ij})$ and $J^{B_{1}B_{2}}
(\mathbf{R}_{ij})$ is equal to $2\theta_{\mathbf{R}}$. But
$J^{A_{1}B_{2}}(\mathbf{R}_{ij})$ and $J^{B_{1}A_{2}}
(\mathbf{R}_{ij})$ have different phase factors
$4\theta_{\mathbf{R}}$ and $0$ respectively. Also we found that at
large distances, the interlayer RKKY interactions between magnetic
located on same sublattice in BLG decay as $1/R^{4}$
approximately. But the interlayer RKKY interactions between
magnetic impurities located on apposite sublattices,
$J^{A_{1}B_{2}}(\mathbf{R}_{ij})$ and
$J^{B_{1}A_{2}}(\mathbf{R}_{ij})$, show decays close to $1/R^{6}$
and $1/R^{2}$ decays at large impurity distances respectively.
Note that only $J^{B_{1}A_{2}}(\mathbf{R}_{ij})$, similar to the
RKKY interaction in ordinary 2DEG, decays as $1/R^{2}$ at large
impurity distances.

To examine our results we discussed the RKKY interaction in BLG in
the two limiting cases, weak $(\gamma\rightarrow0)$ and strong
$(\gamma\rightarrow\infty)$ interlayer coupling. Our results in
these two limiting cases reduced to RKKY interaction in SLG and to
that in BLG in the two-band approximation respectively.

%\newpage

\newpage

\begin{figure}
\begin{center}
\includegraphics[width=7.5cm,angle=0]{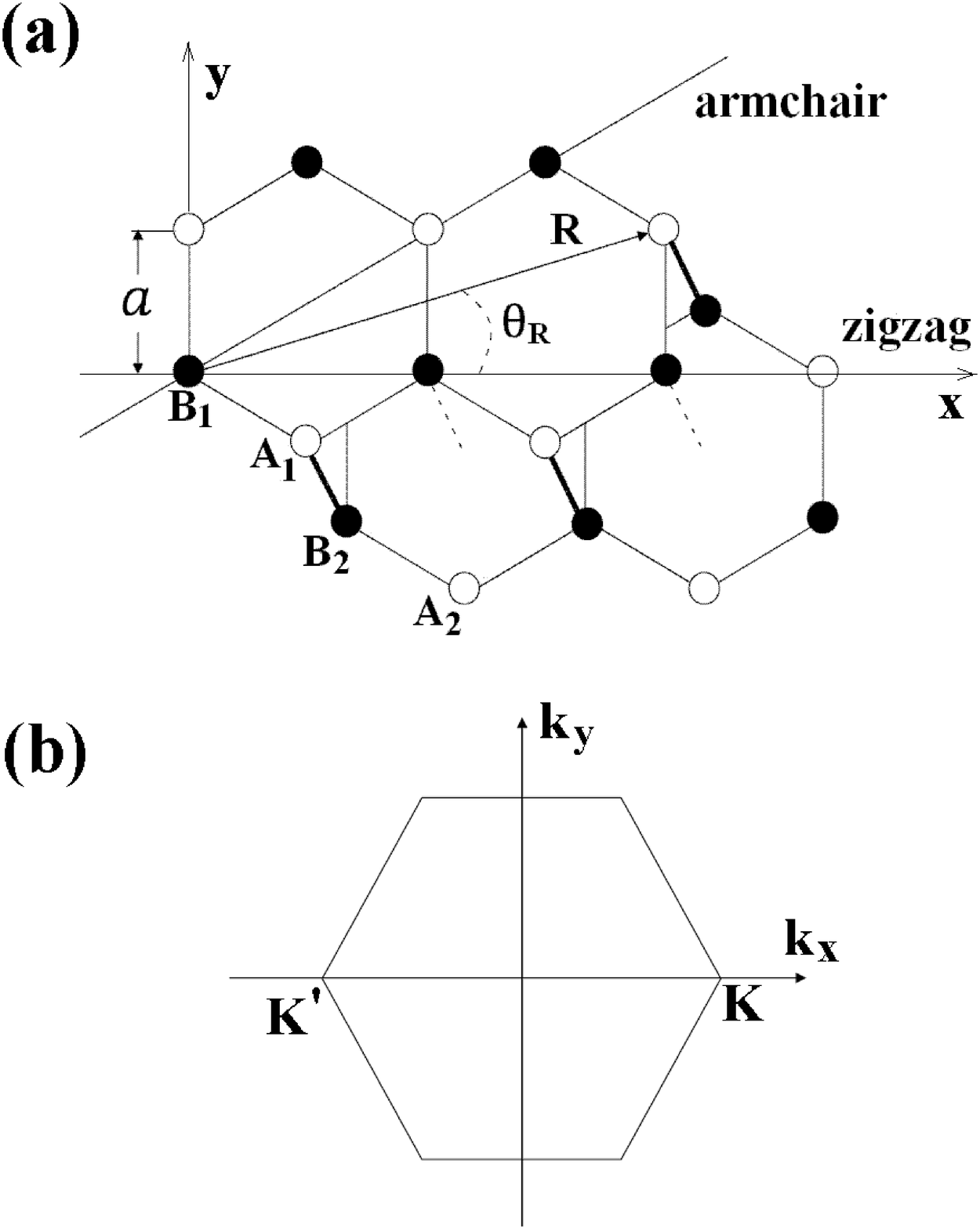}
\caption{(a) Lattice structure of bilayer graphene, (b)
corresponding Brillouin zone, with the Dirac points
$\mathbf{K}=(\frac{4\pi}{3\sqrt{3}a},0)$ and
$\mathbf{K}^{'}=(-\frac{4\pi}{3\sqrt{3}a},0)$
indicated.}\label{fig:01}
\end{center}
\end{figure}

\begin{figure}
\begin{center}
\includegraphics[width=7.5cm,angle=0]{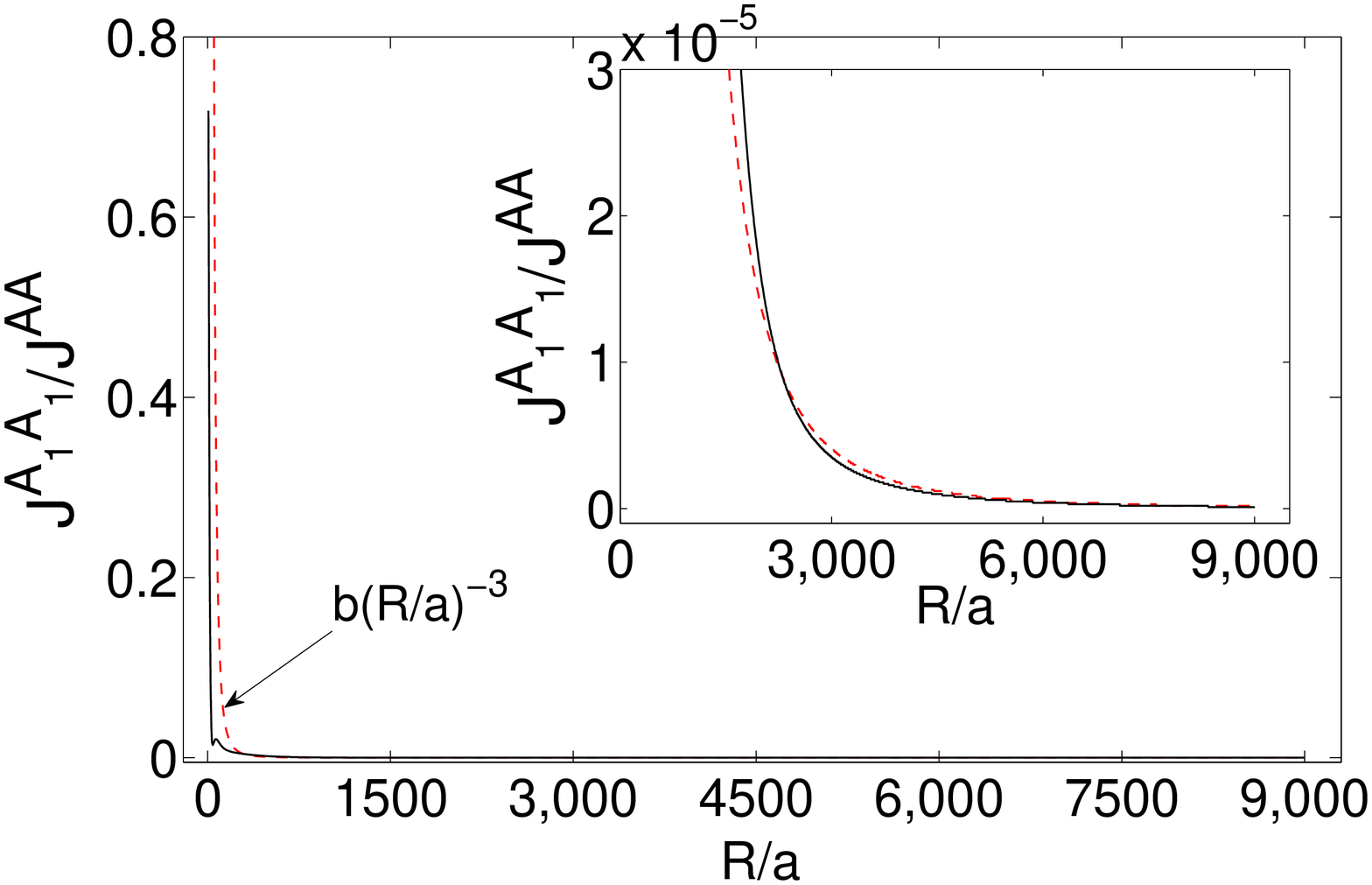}
\caption{ Plot of $J^{A_{1}A_{1}}(\mathbf{R}_{ij})$ in units of
$J^{AA}(\mathbf{R}_{ij})$ (black solid line) fitted to
$\frac{b}{(R/a)^{3}}$ curve (red dashed line) as a function of
$R/a$ where $b=109950$.}\label{fig:02}
\end{center}
\end{figure}
\begin{figure}
\begin{center}
\includegraphics[width=7.5cm,angle=0]{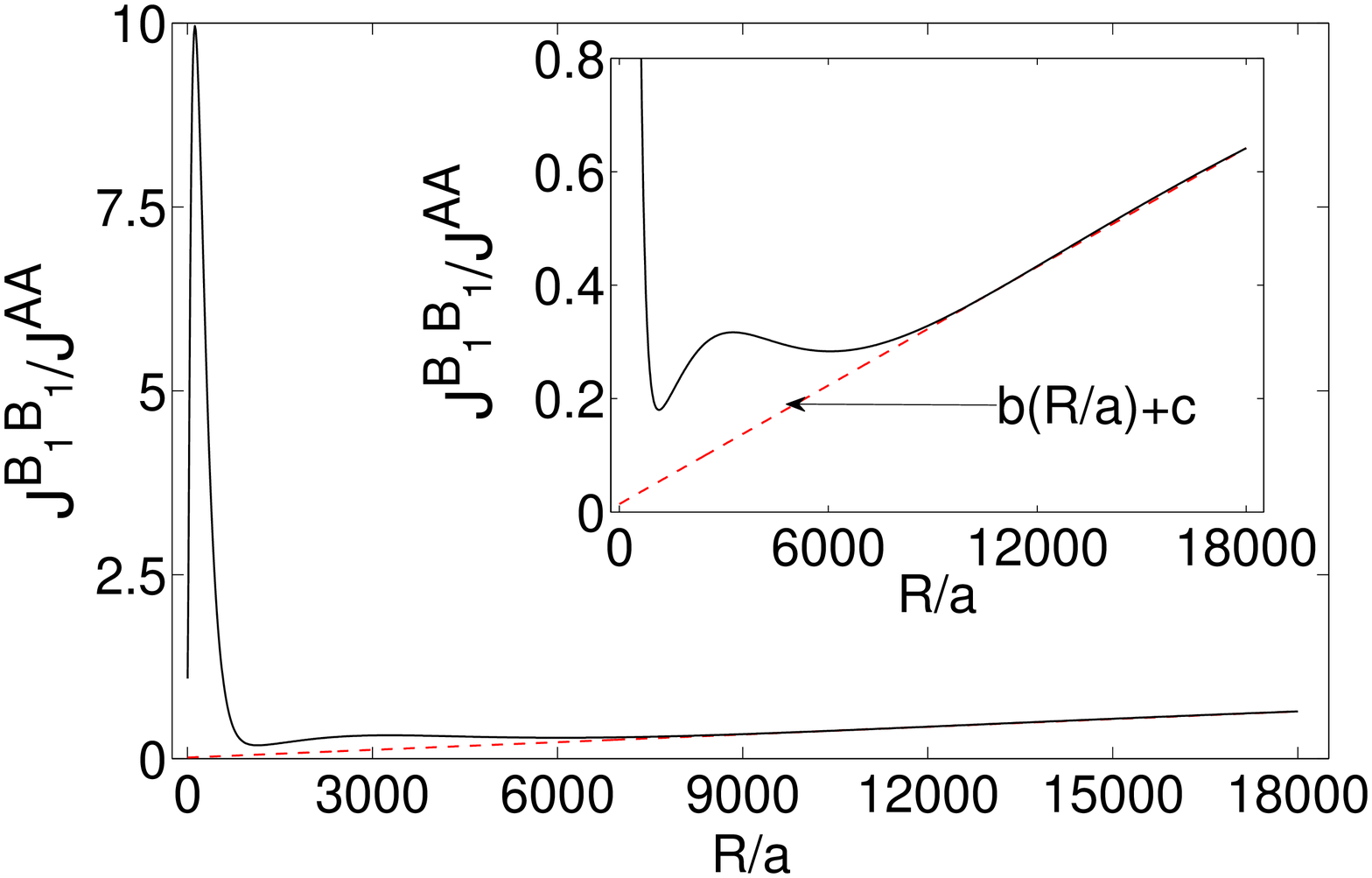}
\caption{ Plot of $J^{B_{1}B_{1}}(\mathbf{R}_{ij})$ in units of
$J^{BB}(\mathbf{R}_{ij})$ (black solid line) fitted to
$b\frac{R}{a}+c$ line (red dashed line) as a function of $R/a$
where $b=0.000035$ and $c=0.0142$.}\label{fig:03}
\end{center}
\end{figure}
\begin{figure}
\begin{center}
\includegraphics[width=7.5cm,angle=0]{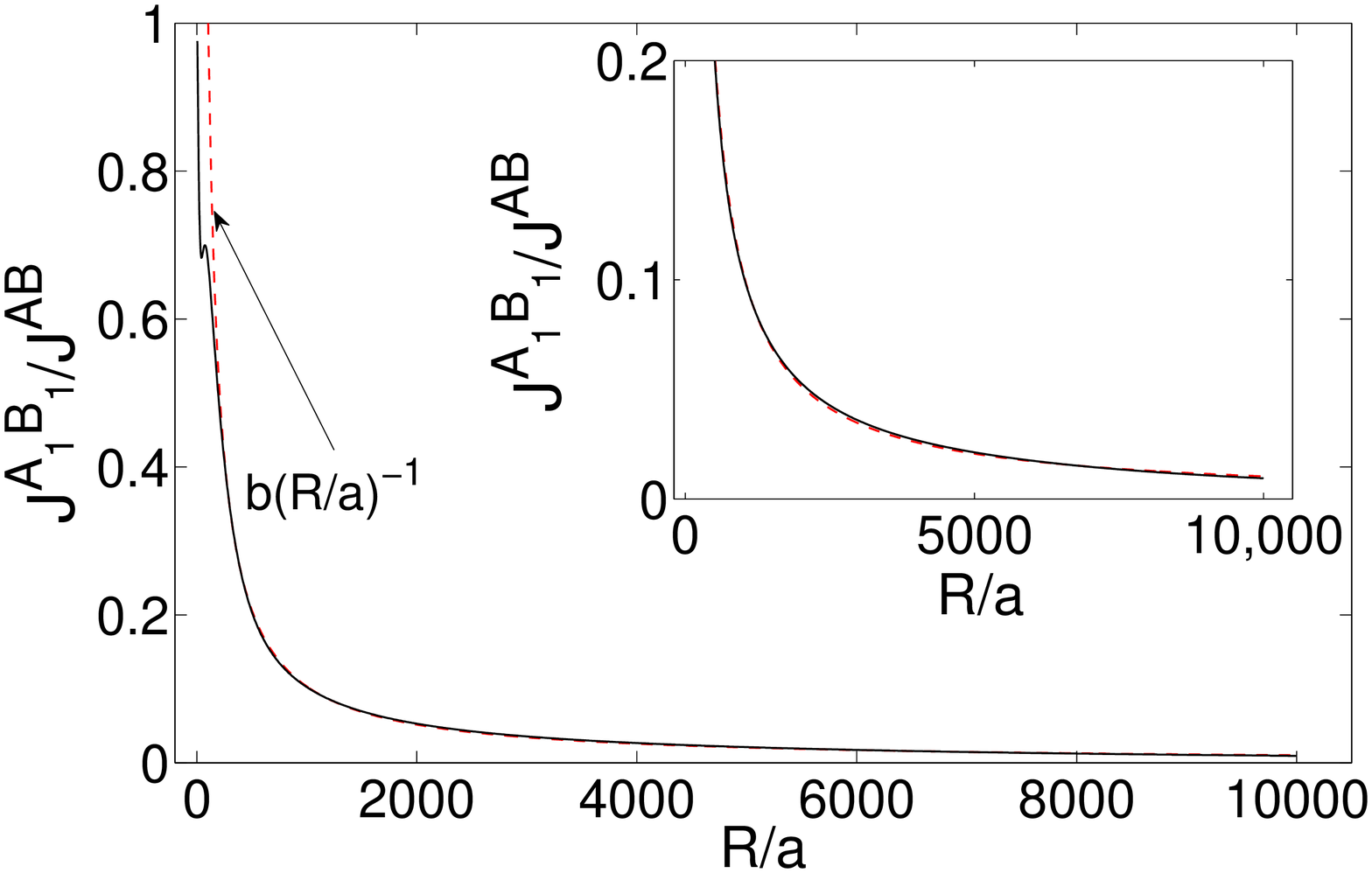}
\caption{ Plot of the $J^{A_{1}B_{1}}(\mathbf{R}_{ij})$ in units
of $J^{AB}(\mathbf{R}_{ij})$ (black solid line) fitted to
$\frac{b}{(R/a)}$ curve (red dashed line) as a function of $R/a$
where $b=103$.
 }\label{fig:04}
\end{center}
\end{figure}
\begin{figure}
\begin{center}
\includegraphics[width=7.5cm,angle=0]{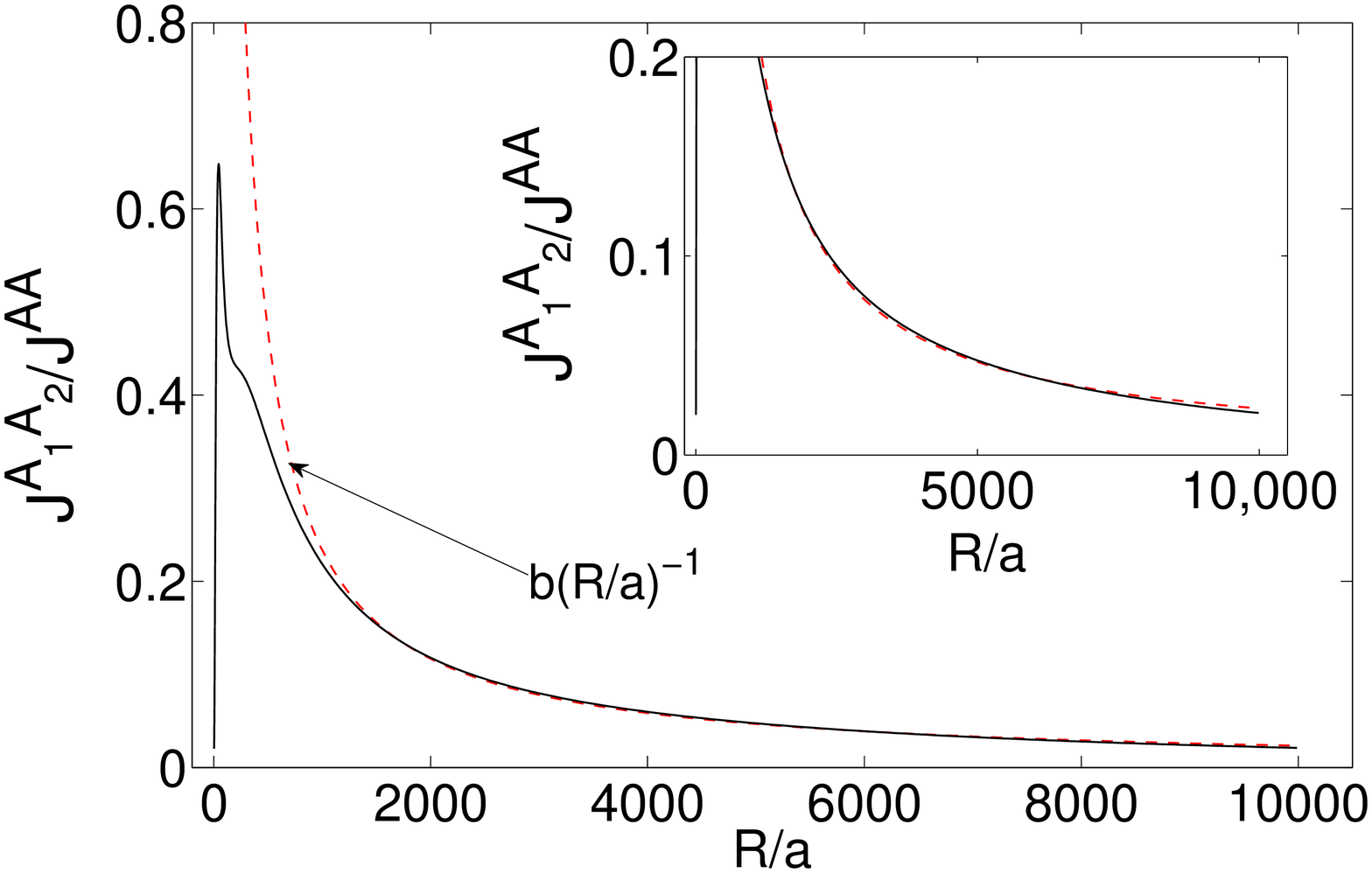}
\caption{ Plot of the $J^{A_{1}A_{2}}(\mathbf{R}_{ij})$ in units
of $J^{AA}(\mathbf{R}_{ij})$ for $\mathbf{R}_{ij}$ along the
armchair direction (black solid line) fitted to $\frac{b}{(R/a)}$
curve (red dashed line) as a function of $R/a$ where
$b=234$.}\label{fig:05}
\end{center}
\end{figure}
\begin{figure}
\begin{center}
\includegraphics[width=7.5cm,angle=0]{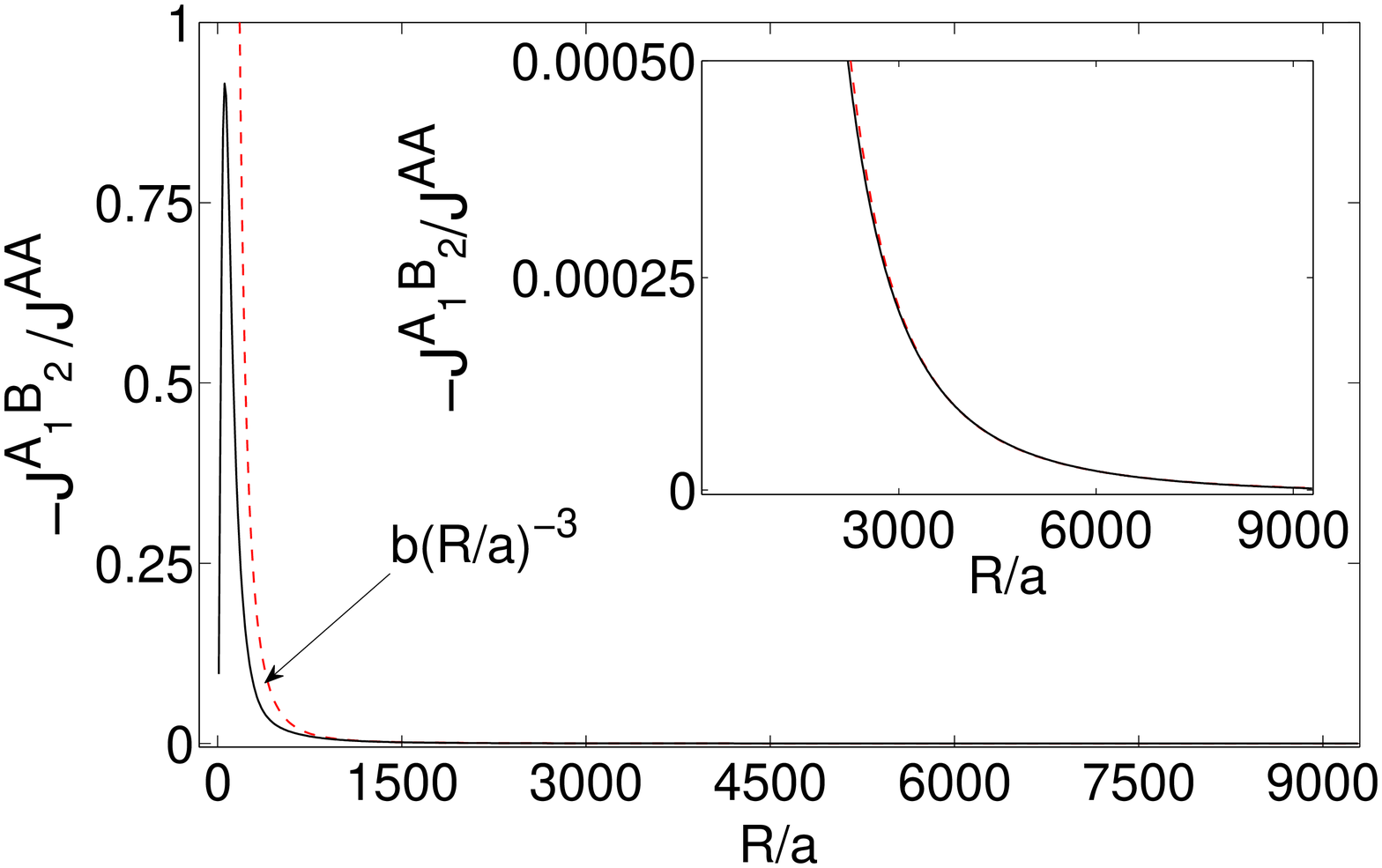}
\caption{ Plot of the $J^{A_{1}B_{2}}(\mathbf{R}_{ij})$ in units
of $J^{AA}(\mathbf{R}_{ij})$ for $\mathbf{R}_{ij}$ along the
armchair direction (black solid line) fitted $\frac{b}{(R/a)^{3}}$
curve (red dashed line) as a function of $R/a$ where
$b=5813000$.}\label{fig:06}
\end{center}
\end{figure}
\begin{figure}
\begin{center}
\includegraphics[width=7.5cm,angle=0]{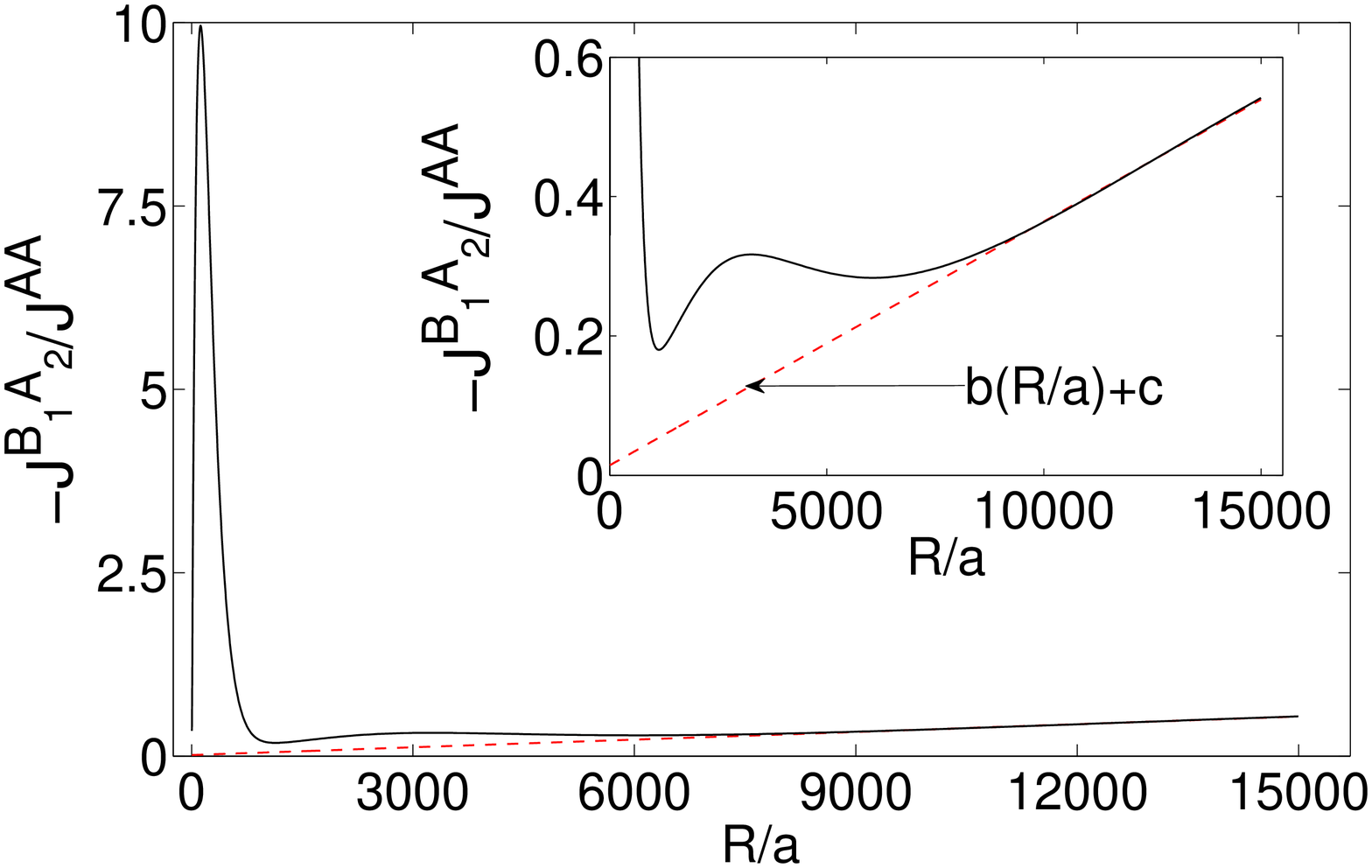}
\caption{ Plot of the $J^{B_{1}A_{2}}(\mathbf{R}_{ij})$ in units
of $J^{AA}(\mathbf{R}_{ij})$ for $\mathbf{R}_{ij}$ along the
armchair direction (black solid line) fitted to $b\frac{R}{a}+c$
line (red dashed line) as a function of $R/a$ where $b=0.000035$
and $c=0.0142$.} \label{fig:07}
\end{center}
\end{figure}


\begin{thebibliography}{a}
\bibitem{Integ1} The integrals in Eqs. \ref{eq:17} and \ref{eq:18} could be solved by using the mathematical identity
for integral of Bessel functions\cite{Kogan,Prudnikov} which are
$\int_{0}^{\infty}dy [\int_{0}^{\infty}xdx J_{0}(x)
e^{-xy}]^{2}=\frac{\pi}{16}$.
\bibitem{Integ2} By using the mathematical identity
for integral of Bessel functions\cite{Kogan,Watson} one can show
that $\int_{0}^{\infty}dy [\int_{0}^{\infty}xdx J_{0}(x)
e^{-x^{2}y}]^{2}=\frac{1}{2}$.
\end{thebibliography}
\end{document}